\documentclass[preprint2]{aastex}

\newcommand{\eh}[1]{\,\mathrm{#1}}

\newcommand{\ttt}[1]{\times10^{#1}}

\newcommand{\dg}{^{\circ}}

\renewcommand{\epsilon}{\varepsilon}

\shorttitle{MAGIC Upper Limits G65.1+0.6}
\shortauthors{Aleksi\'c et al.}

\begin{document}


\title{MAGIC Upper Limits for two Milagro-detected,\\
  Bright Fermi Sources in the Region of SNR G65.1+0.6}


%
\author{
J.~Aleksi\'c\altaffilmark{a},
L.~A.~Antonelli\altaffilmark{b},
P.~Antoranz\altaffilmark{c},
M.~Backes\altaffilmark{d},
J.~A.~Barrio\altaffilmark{e},
D.~Bastieri\altaffilmark{f},
J.~Becerra Gonz\'alez\altaffilmark{g,}\altaffilmark{h},
W.~Bednarek\altaffilmark{i},
A.~Berdyugin\altaffilmark{j},
K.~Berger\altaffilmark{g}
% ...
}
\altaffiltext{a} {IFAE, Edifici Cn., Campus UAB, E-08193 Bellaterra, Spain}
\altaffiltext{b} {INAF National Institute for Astrophysics, I-00136 Rome, Italy}
\altaffiltext{c} {Universit\`a  di Siena, and INFN Pisa, I-53100 Siena, Italy}
\altaffiltext{d} {Technische Universit\"at Dortmund, D-44221 Dortmund, Germany}
\altaffiltext{e} {Universidad Complutense, E-28040 Madrid, Spain}
\altaffiltext{f} {Universit\`a di Padova and INFN, I-35131 Padova, Italy}
\altaffiltext{g} {Inst. de Astrof\'{\i}sica de Canarias, E-38200 La Laguna, Tenerife, Spain}
\altaffiltext{h} {Depto. de Astrofisica, Universidad, E-38206 La Laguna, Tenerife, Spain}
\altaffiltext{i} {University of \L\'od\'z, PL-90236 Lodz, Poland}
\altaffiltext{j} {Tuorla Observatory, University of Turku, FI-21500 Piikki\"o, Finland}
% ...

\email{klepser@ifae.es, decea@ieec.uab.es}



\begin{abstract}
We report on the observation of the region around supernova remnant G65.1+0.6 with the
stand-alone MAGIC-I telescope. This region hosts the two bright GeV gamma-ray sources 1FGL
J1954.3+2836 and 1FGL J1958.6+2845. They are identified as GeV pulsars
and both have a possible counterpart detected at about $35\eh{TeV}$ by the Milagro observatory. MAGIC collected 25.5 hours of good quality data,
and found no significant emission in the range around $1\eh{TeV}$.
We therefore report differential flux upper limits, assuming
the emission to be point-like ($\leq 0.1$\degr) or within a radius of
0.3\degr. In the point-like scenario, the flux limits around $1\eh{TeV}$ are at the
level of 3\% and 2\% of the Crab Nebula flux, for the two sources respectively.
This implies that the Milagro emission is either
extended over a much larger area than our point spread function, or it must be peaked
at energies beyond $1\eh{TeV}$, resulting in a photon index harder than 2.2 in
the TeV band.
\end{abstract}





\section{Introduction}

In February 2009, the Fermi collaboration published a list of the most
significant gamma-ray sources above $100\eh{MeV}$, detected by the large area
telescope (LAT) within 3 months of observation \citep{abd09b}. The energy spectra
of these sources often extend to several GeV, where at some point the steeply
falling flux levels are too low to be detected by the limited detection area of a
satellite instrument.
Many of the LAT sources are hosted by our galaxy, and
34 of those are within the field
of view of the Milagro gamma-ray observatory, which was located near Los
Alamos, New Mexico (\citet{abd09a} and references within). 
The Milagro collaboration, therefore, reinvestigated their previously
gained skymap to look for counterparts to these GeV sources \citep{abd09a}. 
The sensitivity of the instrument peaks in the energy range $10 - 50\eh{TeV}$,
although it ultimately depends on the energy spectrum and the declination of the source.
Due to the reduced trial factor, 14 new Milagro sources could be claimed
with confidence levels above $3\,\sigma$.

Two of these Milagro-detected Fermi bright sources are in
the vicinity of G65.1+0.6, a faint supernova remnant (SNR) first reported by
\citet{lan90}. \citet{tia06} suggested its distance to be $9.2\eh{kpc}$ and a
Sedov age of $40-140\eh{kyr}$. An association to the radio pulsar PSR
J1957+2831 was suggested by \citet{lor98}.

The gamma-ray emission in the region of G65.1+0.6 was first detected 
by the COS-B satellite \citep{swa81} as 2CG065+00, and later confirmed by the
EGRET satellite (3EG J1958+2909) in \citet{har99}, where a possible extension
or multiple sources were denoted. As of now, the two sources could be detected
as two individual sources by the LAT as 1FGL J1954.3+2836 and 1FGL J1958.6+2845, as reported
in the first year catalog of Fermi sources \citep{abd10b}. They were
analysed and reported as gamma-ray pulsars found through blind search in
\citet{saz10}, \citet{abd09c} and \citet{abd10a}. Their periods ($290\eh{ms}$, $92.7\eh{ms}$),
spin-down luminosities ($104.8\ttt{34}\eh{ergs} \eh{s^{-1}}$, $33.9\ttt{34}\eh{ergs}
\eh{s^{-1}}$) and energy cutoffs ($2.9\eh{GeV}$, $1.2\eh{GeV}$), but also the
characteristic magnetic fields and ages ($69.5\eh{kyrs}$,  $21\eh{kyrs}$)
lie in the average range of
all Fermi pulsars.
In the following, we will use the names J1954
and J1958 for these 1-year catalog sources, and J1954$_0$/J1958$_0$ to
specifically refer to the 3-month bright source list positions.

The detections by Milagro revealed significances of $4.3\,\sigma$ for
J1954$_0$ and $4.0\,\sigma$ for J1958$_0$ \citep{abd09a}. Flux values are stated
for a characteristic median energy of $35\eh{TeV}$. The angular
resolution of Milagro is about $0.4 - 1.0\dg$, so these values can be
expected to hold also
for the 1-year catalog positions of the sources, which are offset by $\leq 0.1\dg$.
Since gamma-ray
pulsars typically have energy cutoffs as low as several GeV, the Milagro
signals, if real, can be expected not to be caused directly by the pulsars,
but possibly by associated objects, such as a pulsar wind nebula (PWN) or,
in the case of J1954, the shell of the SNR, which surrounds the pulsar.
Gamma rays at TeV energies may also be produced in an interaction of the shell
with a possibly coincident molecular cloud.
This cloud may be located at the position of the infra-red source IRAS
19520+2759, which has associated CO line, H$_{2}$O and OH maser
emission at a similar distance as the SNR \citep{arq87}.



The MAGIC telescopes use the Cherenkov imaging technique and are located on
the Canary Island of La Palma (28.8\degr N, 17.8\degr W, 2220 m a.s.l.). It is
the instrument
with the lowest energy threshold 
among all Cherenkov telescopes. In single-telescope observations, as presented
here, the nominal threshold in low zenith
angle observations is $60\eh{GeV}$ \citep{alb08a}. It is therefore the instrument of choice
to connect the upper ends of the Fermi spectra, which typically end at some
tens of GeV, with the detections provided by
Milagro at some tens of TeV.
To bridge these spectra in the TeV range, and identify possible object
associations for the Milagro
signals were the main motivations for our investigation.

\section{Observations} \label{sec_obs}

Motivated by the fact that J1954$_0$ was \textit{not} marked as a pulsar in the
initial Fermi
Bright Source List, we observed J1954$_0$ as a main target in July and
August 2009. The observations were carried out in false source tracking
(wobble) mode \citep{fom94}, which yielded two datasets with offsets of
$\pm$0.4\degr\ in RA from this source, see Figure~\ref{fig1}. The wobble position was altered every
$20\eh{min}$, and the data were taken at zenith angles between 0
and 43 degrees. At the time,
the second MAGIC telescope was still under comissioning, so the analysis
presented here used only the data from the stand-alone MAGIC-I telescope.

Quality selection
cuts were applied to the event rate, the spread of hardware-sensitive shower
image parameters, and few parameters that characterize the transparency of the
atmosphere, such as the sky temperature and humidity.
After all data selection cuts, 25.5 hours of high quality data were left for
the analysis of J1954.

Besides this, we took advantage of the fact that J1958 is in the field of view
of one of the two wobble positions. It could be analysed in a specific way
described in the next section, yielding 13.8 hours of effective
observation time.


\section{Analysis and Results} \label{sec_ana}

The data were analysed in the MARS analysis framework \citep{mor09}, which is the standard
software used in the analysis of MAGIC data. After the air shower images of the
photomultiplier tube camera
are calibrated, and times and charges of each pixel are extracted, a three-stage image
cleaning is applied to filter out uncorrelated noise from the data acquisition
electronics and the night sky background \citep{ali09}. Shower image parameters are calculated,
and the Random Forest method \citep{alb08b} is used to derive estimators for the shower
direction, its energy, and its likeliness to be of hadronic origin.

\begin{figure}
\plotone{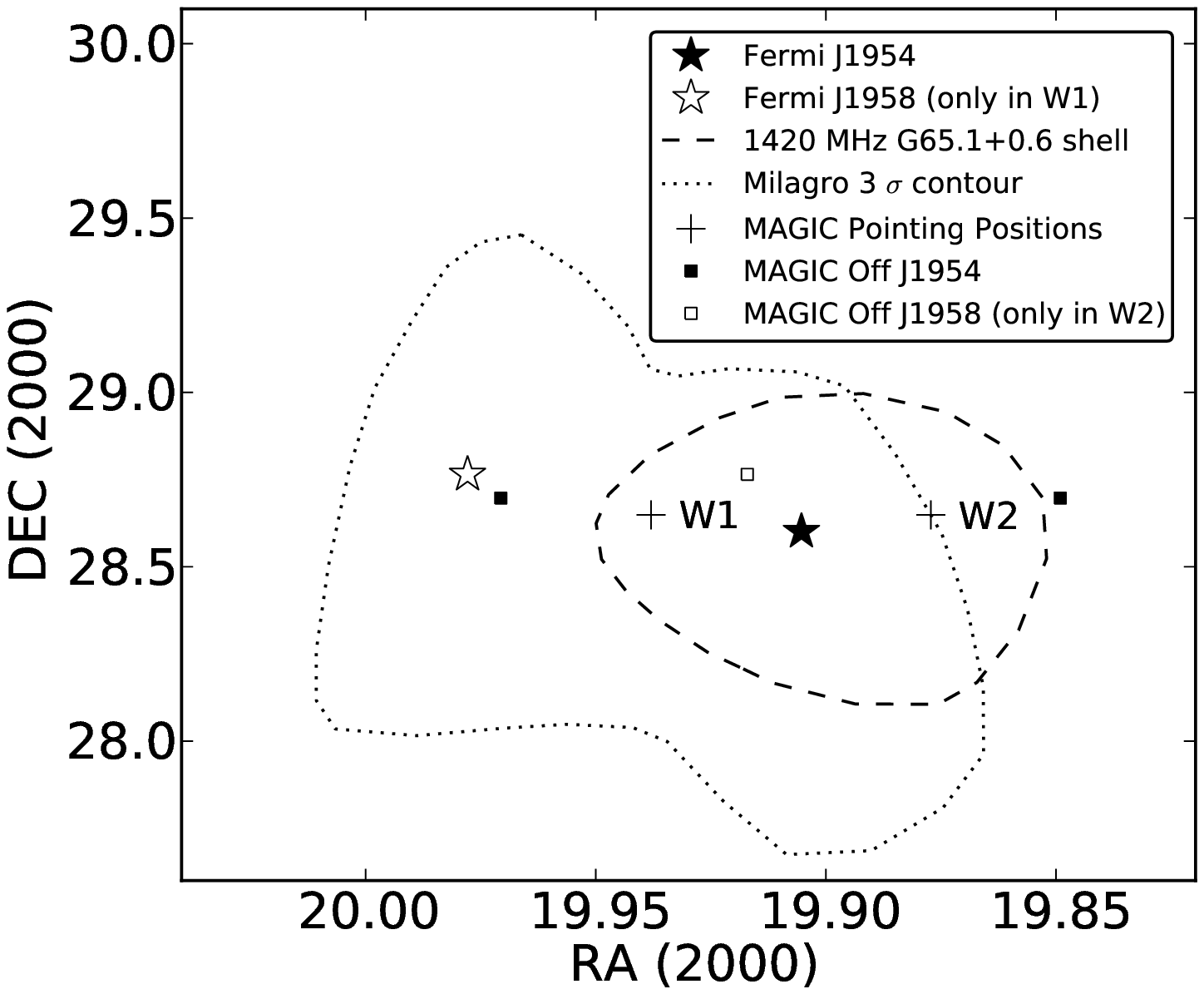}
\caption{Observation setup for the two Fermi sources J1954 and J1958 in the
context of SNR G65.1+0.6 and a Milagro significance contour. 
J1958 appears
only in one wobble position (W1), so the OFF data is taken from the other wobble
sample, using the same position relative to the pointing direction. The
outline of the remnant is taken from the radio map in \citet{lan90}. The
extension of the Milagro significance contour \citep{abd09a} is compatible
with their point spread function.  
\label{fig1}}
\end{figure}

The observational setup described in the previous section requires a different
analysis treatment of both sources. The main target of the observation, J1954,
can be analysed using standard wobble analysis procedures.
This means that the photon flux from the source is compared to the one on the
opposite side of the camera (\textit{anti}-source, solid squares in Figure~\ref{fig1}).
In this way, exposure inhomogeneities that can arise from imperfections in the
photo multiplier tubes, trigger electronics or the signal transmission, cancel out, because both wobble
positions were equally populated.

In the case of J1958, a wobble analysis, using only one of the two
wobble positions, is not guaranteed to
cancel out these  inhomogeneities.
Therefore, the analysis was done in ON/OFF manner, using the near wobble sample as
ON-source data and the far wobble sample as OFF-source data (hollow squares in
Figure~\ref{fig1}). Having the OFF-source at the same position in relative camera coordinates as the source in
the ON sample, the exposure inhomogeneities cancel out.

To test the presence of a gamma-ray signal, the distributions of squared angular distances
($\theta^2$) between photon directions and the source positions were used. In
these $\theta^2$-plots, a signal is expected to produce an excess where $\theta^2$
approaches zero. However, the integrals over the expected signal
regions of the $\theta^2$ distributions agree, for both sources, well with the
corresponding integrals done with respect to the OFF regions. Furthermore, the
shapes of the ON and OFF distributions agree well with each other, and are sufficiently flat to exclude
the unlikely case of an emission occuring by chance at both ON and OFF
locations at a similar flux level, supressing the significance of the $\theta^2$
comparison. Such a coincidence was also excluded by cross-checking the
analysis using several OFF regions, and by thoroughly investigating the skymaps with different
background estimation algorithms. Consequently, we report the absence
of a significant signal for both sources.


%

Besides this analysis, which takes advantage of a-priori defined source
locations, also a 2.8\degr$\times$2.0\degr\, skymap of the area was
investigated at different energy ranges. The trial factors implied by these
searches were typically between 120 and 260, depending on the PSF, which is
smaller for high energies. Taking into account these trials,
no significant signal was found at any energy.

We convert our data into three differential flux upper
limits for each source. To do so, the data was divided into three bins of
estimated energy, delimited by $120\eh{GeV}$, $375\eh{GeV}$, $2.8\eh{TeV}$ and
$12\eh{TeV}$. For each bin, an \textit{event number} upper limit is
calculated from the above-mentioned $\theta^2$-plots, at $95\%$ confidence
level (c.l.) after \citet{rol05}, and assuming an efficiency systematic error of
$30\%$ \citep{alb08a}. For each limit, a power law energy spectrum with a
photon index of $-2$ is assumed both for the simulation of the effective area,
and the conversion of event number upper limits to \textit{flux}
upper limits. The influence of this
photon index is minor, since the energy ranges are sufficiently small.

Finally, since the data can only be selected by \textit{estimated} energy, a
Monte Carlo simulation (MC) is used to estimate the median \textit{true}
energy of the data that remains after all cuts.


This analysis assumes a source extension similar or smaller than the point
spread function (PSF) of MAGIC. In an identically conducted analysis of
contemporary Crab Nebula data the width of this PSF (defined as the Sigma of a
two dimensional Gaussian function) was found to be about 0.08\degr. Since the Milagro source may be extended,
a second analysis was done, assuming an extension of 0.3\degr\ instead. Since this is done simply by increasing the signal integration
radius, more background events are included, which leads to higher upper limits.

\section{Discussion} \label{sec_res}

The derived 95\% c.l. flux upper limits are summarized in
Table~\ref{tbl-1}.
Figures~\ref{fig2} and~\ref{fig3} display them in the
context of the published satellite data and the Milagro flux estimation. The
differences between the limits of J1954 and J1958 are all compatible with the
statistical fluctuations that can be expected for 95\% upper limits.
Around $1\eh{TeV}$, where MAGIC is most sensitive, the flux is limited to 3\% of the Crab
Nebula flux for J1954 and 2\% for J1958. Assuming that the Milagro emissions originate
from objects spacially coinciding with the Fermi pulsars within our PSF,
the photon index in the energy range of $1$ to $35\eh{TeV}$ must be harder
than 2.2 for J1954, and 2.1 for J1958. The spectral energy distribution is thus likely to peak at energies
in excess of $1\eh{TeV}$.

If an extension up to 0.3\degr\ is assumed, 
the corresponding flux limits in Crab Nebula units are 14\% for J1954 and 3\% for J1958.
In this extended case, the photon indices are
limited to $\leq 2.6$ and $\leq 2.2$, respectively.
It shall be noted that the biggest TeV pulsar wind nebulae have sizes of few
tens of parsecs, which at the distance of G65.1+0.6 would be within these
0.3\degr.

Bringing together the existing flux data and our upper limits, we conclude that the most likely scenario
to explain the gamma ray production measured by Milagro might be the
existence of two PWN, associated with the Fermi pulsars. With the ages of the
pulsars being $69.5\eh{kyrs}$  and $21\eh{kyrs}$, it is reasonable to expect
an inverse compton component that dominates their energy outflows and may
additionally be extended \citep{jag09, tan09}. Such old, extended PWN
are common TeV gamma-ray sources and frequently have an emission spectrum that peaks at TeV energies
or above. 

\section{Summary} \label{sec_sum}

\begin{figure}
\plotone{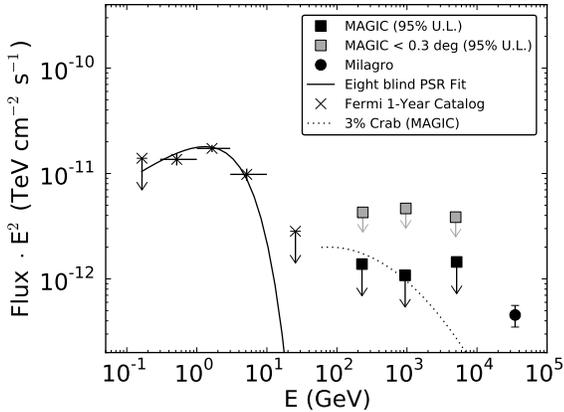}
\caption{Compilation of flux measurements and upper limits for 1FGL
J1954.3+2836 from Fermi \citep{saz10, abd10b}, MAGIC and Milagro
\citep{abd09a}. The 3\% fraction of the MAGIC Crab spectrum \citep{alb08a} is shown for
comparison.\label{fig2}}
\end{figure}

\begin{figure}
\plotone{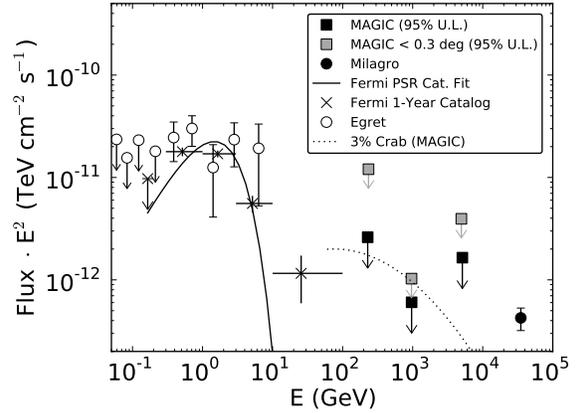}
\caption{Compilation of flux measurements and upper limits for 1FGL
J1958.6+2845 from EGRET \citep{har99}, Fermi \citep{abd10a, abd10b}, MAGIC and Milagro
\citep{abd09a}. The 3\% fraction of the MAGIC Crab spectrum \citep{alb08a} is shown for
comparison.\label{fig3}}
\end{figure}

We took 25.5 hours of good quality data in the area of the faint supernova remnant
G65.1+0.6. In that region, the Milagro collaboration reported the emission of
gamma rays with median energy of $35\eh{TeV}$ in the vicinity of
the two GeV Fermi pulsars 1FGL J1954.3+2836 and 1FGL J1958.6+2845. Our
observations, which were aimed to locate the source of the Milagro emission,
yielded no
significant gamma-ray signal for these two a-priori known source locations. Also, no 
post-trial significant signal from a skymap of the area could be established. We extracted three
differential flux upper limits for each source, assuming two
different extension radii. They are summarized in Figures~\ref{fig2}
and~\ref{fig3}.

Assuming that the two 4 and $4.3 \sigma$ detections of Milagro are not statistical
fluctuations, but real signals, the flux upper limits support the scenario in
which the multi-TeV emission measured by Milagro is caused by a different
mechanism or object than the Fermi emission. Given the ages of the pulsars and the SNR,
the existence of two very old pulsar wind nebulae, powered by the two GeV pulsars,
seems very likely.

\acknowledgments
We would like to thank the Instituto de Astrofisica de 
Canarias for the excellent working conditions at the 
Observatorio del Roque de los Muchachos in La Palma. 
The support of the German BMBF and MPG, the Italian INFN,
the Swiss National Fund SNF, and the Spanish MICINN is gratefully
acknowledged. 
This work was also supported by the Polish MNiSzW Grant N N203 390834, 
by the YIP of the Helmholtz Gemeinschaft, and by grant DO02-353
of the the Bulgarian National Science Fund.






\appendix

\clearpage



\clearpage

\clearpage

\begin{deluxetable}{ccrrrr}
\tabletypesize{\scriptsize}
\rotate
\tablecaption{Differential Upper limits\label{tbl-1}}
\tablewidth{0pt}
\tablehead{
\colhead{Source Name} &
\colhead{Assumed Extension} &
\colhead{$E_{\mathrm{med}}$} &
\colhead{Significance} &
\colhead{$F_{95\%}$} &
\colhead{$F_{95\%}E^2$} \\
\colhead{} &
\colhead{(deg)} &
\colhead{(GeV)} &
\colhead{($\sigma$)} &
\colhead{(10$^{-12}\,$TeV$^{-1}$cm$^{-2}$s$^{-1}$)} &
\colhead{(10$^{-12}\,$TeVcm$^{-2}$s$^{-1}$)} 
}
\startdata
1FGL J1954.3+2836   & $\leq 0.08\dg$ & 228  & -1.8 & 27    & 1.38 \\
                    & & 942  & +1.1 & 1.22  & 1.08 \\
                    & & 5123 & +1.9 & 0.055 & 1.45 \\
                    & $\leq 0.3\dg$  & 234  & -1.3 & 78    & 4.3\\
                    & & 963  & +2.6  & 5.0   & 4.66\\
                    & & 4956 & +2.0  & 0.157 & 3.85\\
1FGL J1958.6+2845   & $\leq 0.08\dg$ & 228  & -1.5 & 50    & 2.60 \\
                    & & 966  & -0.9 & 0.65  & 0.60 \\
                    & & 5123 & +0.9  & 0.063 & 1.65 \\
                    & $\leq 0.3\dg$  & 234  & -0.3 & 220   & 12.0\\
                    & & 963  & -1.9 & 1.11  & 1.03\\
                    & & 4956 & +0.8 & 0.161 & 3.9\\
\enddata
\end{deluxetable}

\end{document}